# Scenic Routes in $R^d$

**Kamalakar Karlapalem**  KAMAL@IIIT.AC.IN
*Data Science and Analytics Centre*
*International Institute of Information Technology*
*1 Prof. C. R. Rao Road, Gachibowli, Hyderabad*
*India.*

**Abstract**

In this work, we introduce the problem of scenic routes among points in $R^d$. The key development is the nature of the problem in terms of both defining the concept of scenic points and scenic routes and then coming up with algorithms that meet different criteria for the generated scenic routes. The scenic routes problem provides a visual trajectory for a user to comprehend the layout of high-dimensional points. The nature of this trajectory and the visual layout of the points have applications in comprehending the results of machine learning supervised and unsupervised learning techniques. We study the problem in 2D and 3D (with two color points) before exploring the issues in $R^d$. The red/blue points in our examples could be to be in a class or not to be in a class. The applications could include landscape design to adhere to the scenic beauty of the artifacts on the ground. The generation of equally separated layouts for designing composite hardware where interference could be an issue.

**Keywords:** scenic points, scenic lines, scenic planes, scenic regions, traversal graphs, scenic route, optimization

## 1   Introduction

In the real world, there is a well-understood concept of a scenic point or route. Especially when traveling through the wonders of nature. Humans can instinctively identify scenic spots or routes (a path) well recognized as scenic by other humans. A scenic spot in nature can be a point in hilly terrain looking at the valley with mountain tops, a vast sea with cliffs and beaches, an encompass of dunes across a desert, a vastness of artic ice cover, etc. Given a set of points in $R^d$, which spread across the $R^d$ space, there can be points in $R^d$ from which the closer points satisfy certain scenic criteria. The scenic criterion is a relationship between the scenic and other points. Contiguous scenic points can form a line, curve, plane, region, etc., giving scenic space areas. These areas can intersect, giving smaller areas that provide multiple points, each meeting scenic criteria. To appreciate the comprehensiveness of the scenic routes problem, we introduce examples and specific scenarios from $R^2$, and $R^3$, and then present open research problems for $R^d$. The emphasis of this work is to introduce these ideas incrementally and to present a landscape of open problems that interest bout theorists and applied scientists.

We present in Section 2 the scenic routes problem over points in 2D; in Section 3, extend the 2D problem to weighted points, in section 4 we introduce the scenic route problem in 3D; in Sections 5 and 6, we connect the ideas to scenic routes problem in Rd, and finally in section 7, we present some conclusions.

## 2  Scenic Routes over Points in $R^2$ (2D) Space

In this section, we present the simplest scenic route problem and then evaluate different aspects of the problem and solutions. Consider a 2D coordinate space with a set of red and blue points. We define a scenic point as a point that is equidistant to a red point and a blue point. The set of contiguous scenic points forms a scenic path. The perpendicular bisectors to the line joining a red and blue point form a scenic path between the red and blue points. The scenic perpendicular bisectors between different pairs of red and blue points can intersect, forming a scenic graph of intersection points and edges. A scenic route is a traversal on a scenic graph.

Let the set of red points be $R$, the set of blue points be $B$, and the set of points where the bisectors intersect be $I_P$. Segments of the bisectors between these intersection points form the edge set $E_P$, contributing to the scenic paths' edges. The weight of each edge is the Cartesian distance between its endpoints. The graph over all the intersection points is $G(I_P, E_P)$. Moreover, for a selected route, let the set of points on a route be $S \subseteq I_P$, and the edges within the route be $E \subseteq E_P$. We explain the problem further by using Figure 1. Consider the point configuration containing one red point $\{R_1\}$ and three blue points $\{B_1, B_2, B_3\}$. $\{M_1, M_2, M_3\}$ represent the midpoints of the lines joining them. The green lines are the perpendicular bisectors of these lines. The green lines represent scenic paths. The intersection points of these perpendicular bisectors are $\{IP_1, IP_2, IP_3\}$. Therefore, to obtain a scenic route, we consider the perpendicular bisectors. Further, we use their intersection points to move from one bisector to another. In this case:

$$I_P = \{IP_1, IP_2, IP_3\}$$
$$E_P = \{[IP_1, IP_2], [IP_2, IP_3], [IP_3, IP_1]\}$$

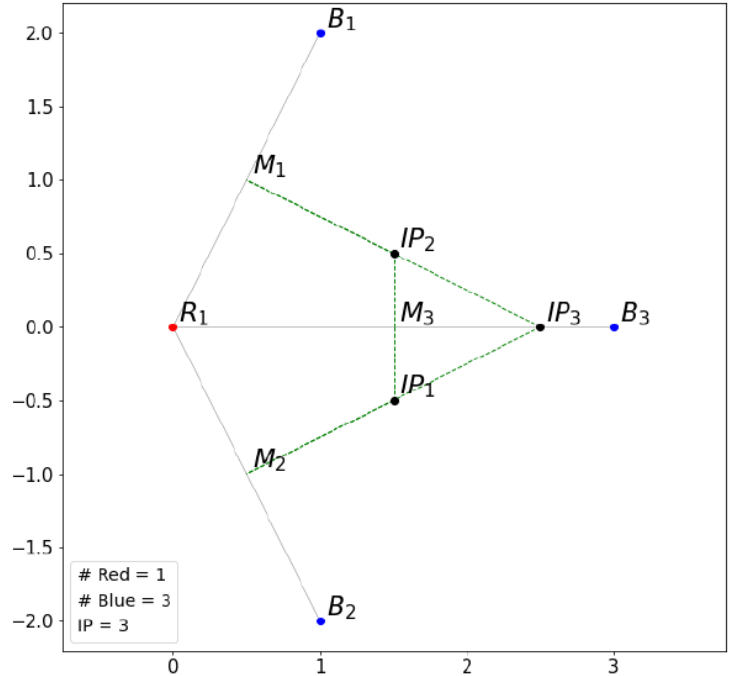

Figure 1: Red/Blue points, bisector intersection points ($I_P$'s), and corresponding scenic graph.

A scenic route has points $\{IP1, IP2, IP3\}$ and edges $\{[IP_1, IP_2], [IP_2, IP_3], [IP_3, IP_1]\}$; any one of these edges provides a complete scenic view of all pairs of red/blue points. The complete triangle of points provides multiple perspectives of the scenic points.

## 2.1 Scenic Routes/Traversals in 2D

A scenic route is defined as a route over $I_P$ via the bisectors (since it is over $I_P$ and we only use bisectors, it is guaranteed to consist of scenic paths). Many such routes with different characteristics are possible. Therefore, there needs to be a characterization of the properties of a scenic route.

A scenic route has the following requirements, decreasing in order of importance:

1) **Only Scenic:** The route must consist of scenic paths. Any non-scenic path is distracting and must be avoided.
2) **Completeness:** Travelling on the route must allow one to view many red-blue pairs. It is preferable for a route to give a view of all red-blue pairs (that is, all $|R| \cdot |B|$ pairs). The ability to view a larger number of red-blue pairs on a scenic route would add to the scenic beauty the route offers. Ideally, all scenic views for all pairs of red-blue points must be covered by a scenic route.
3) **Minimal Edges:** A route must not have many edges. Traveling on a route must allow for long, uninterrupted stretches of scenic points. In other words, a route should have few (not many) direction changes.
4) **Minimal Repeated Edges:** A route must minimize the number of repeated edges. Repeated edges are defined as stretches of bisectors that must be traveled multiple times (repeated) to complete the entire route. Repeated edges come into play to produce a closed path to return to the starting point. Repeated edges within scenic routes offer the same view multiple times. These edges unnecessarily increase the total length of the scenic route without offering any additional scenic views. Hence, they should be minimized. In Fig. 3, the edge *[α, β]* is an example of a long-repeated edge. If a user were to travel on this edge, the user would also need to travel back on the same edge, i.e., repeat the edge.

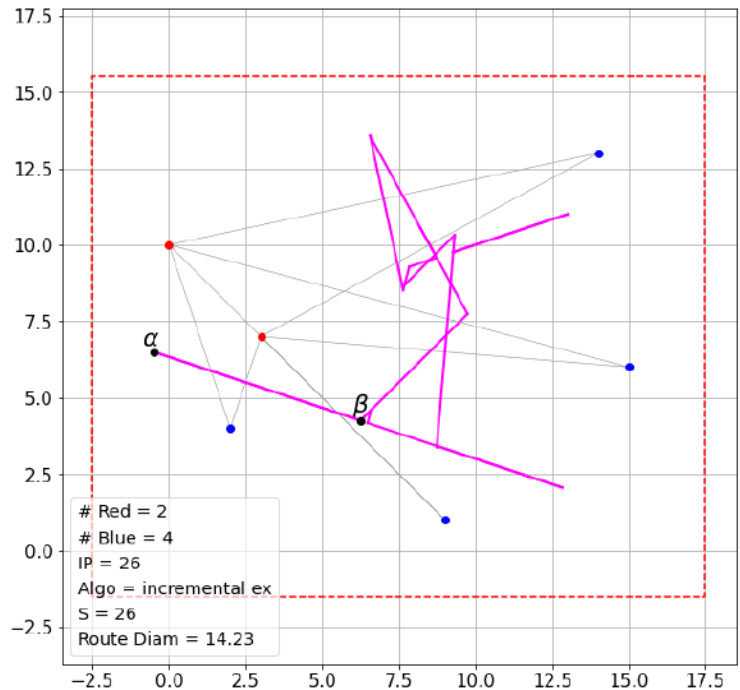

Figure 2: A repeated edge *[α,β]*. ($|R| = 2$, $|B| = 4$, $|I_P| = 26$)

To better understand why the above requirements and the ordering of these requirements generates a preferred scenic route, consider a person walking on a scenic route.
- At all points during travel, the person should have a scenic view available (Req. 1).
- Given that the red-blue points are points of interest, a person would want to view as many (preferably all) red-blue points as possible (Req. 2).
- A person walking on a route would dislike many direction changes and would prefer longer paths to not distract from the scenic beauty available. (Req. 3).
- Finally, a person would not want to traverse the same path repeatedly because repeated edges do not offer any additional novel views, therefore unnecessarily increasing the total distance that needs to be traversed (Req. 4).

A route that fulfills all four abovementioned requirements is a preferred scenic route. A route that fulfills fewer than four requirements is still scenic; however, such a route would not be preferred. For instance, a route may fulfill the first three requirements but include some repeated edges. Moreover, the importance of these requirements and their tradeoffs must be considered when generating scenic routes. Having no repeated edges but viewing a small number of red-blue pairs is worse than having many repeated edges but viewing many red-blue pairs.

**Spatial Restrictions**: The perpendicular bisectors between two pairs of red-blue points can be almost parallel, generating intersection points far from the red-blue pairs. While the view given by such points is still scenic according to our definition, it is not practical. Removing such distant intersection points from the set *IP* makes sense in such cases.

**2.2   Algorithms for Scenic Routes in 2D**

Each algorithm needs graph $G(I_P, E_P)$, where $I_P$ is the set of intersection points of the bisectors and $E_P$ is the set of edges (scenic paths) between these intersection points. The algorithms designed require knowing the shortest path that connects a pair of points. This shortest path is computed using the Floyd-Warshall all-pairs shortest path algorithm. We use the notation $d(a, b)$ to denote the distance between two points taken from the pre-computed results of the Floyd-Warshall algorithm. The algorithms use the pre-calculated output of the All Pairs Shortest Path (APSP) Algorithm on $G(I_P, E_P)$.

Following algorithms that meet various criteria are designed:
- **Min-Max Hull Algorithm:** The MinMax Hull algorithm aims to get a convex scenic route with no repeated edges that satisfy the distance bound. Convex scenic routes perform well on two scenic requirements: they have long, uninterrupted views (Req. 3) and have no repeated edges (Req. 4). The blue route in Figure 3 included extra edges (due to lack of direct connectivity between two intersection points) provides the scenic route for these four red and four blue points.
- **Densest Line Algorithm:** The Densest Line algorithm aims to identify long, straight, uninterrupted scenic paths that reduce directional changes and maximize the number of views. It would make sense to connect the endpoints of the long bisectors using a hull so that once a user reaches the end of a bisector, the user can shift to another bisector using the hull instead of having to backtrack on the same bisector. However, a route consisting solely of long bisectors would perform poorly on the repeated edges requirement (Req. 4). Therefore, there need to be some connecting paths between the long bisectors that allow users to view other red-blue pairs without traversing back on the same long bisector. It would make sense to connect the endpoints of the long bisectors using a hull so that once a user reaches the end of a bisector, the user can shift to another bisector using the hull instead of having to backtrack on the same bisector. In particular, we choose alpha shapes [Edelsbrunner et al.(1983)] as our hull algorithm.

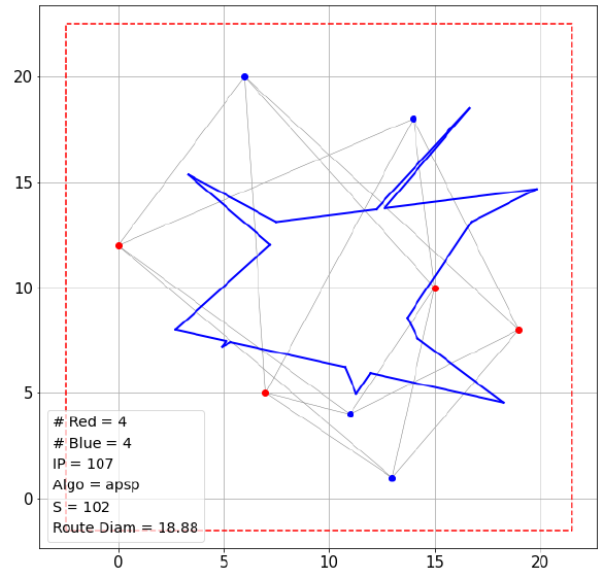

Figure 3: Min-Max Hull, bounded by a bounding box. *(|R| = 4, |B| = 4, |I_P| = 107)*

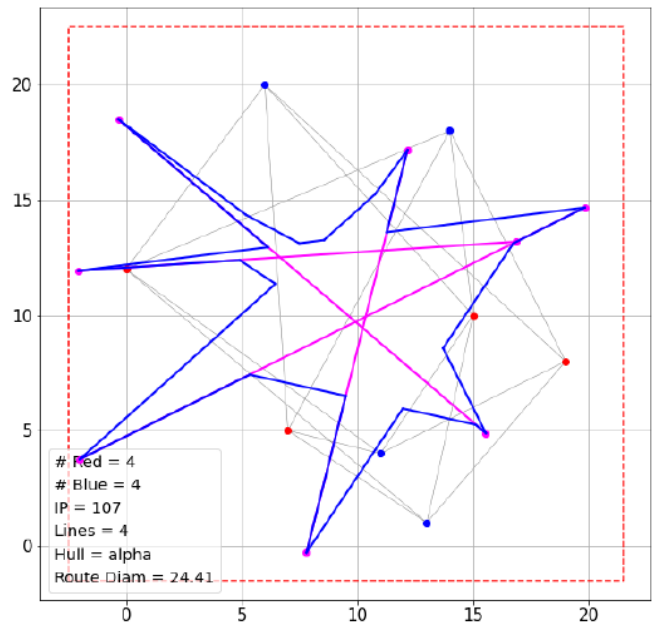

Figure 4. Densest lines (in pink) with alpha shape hull (hull in a dark blue) scenic route

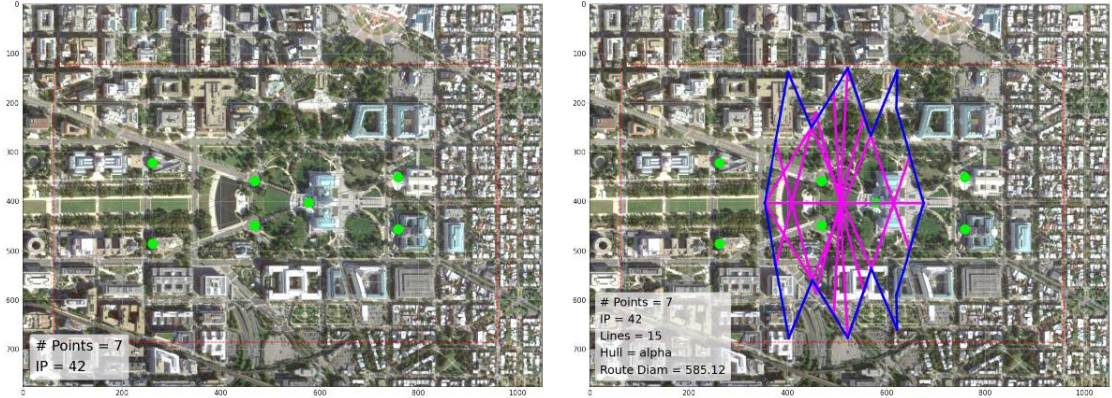
Figure 5: Green dots are landmarks, and the cyan blue route is scenic, covering the landmarks.

## 2.3 Summary

The scenic routes problem with red/blue points is challenging, especially as the number of intersection points increases as the number of red/blue points increases. If one were to look for optimal criteria such as completeness (all pairs of red/blue points must be visited), the total distance traveled should be minimum and with fewer repeated edges. In the min-max hull and densest line algorithm, neither completeness nor optimal route is guaranteed. Further, whether an optimal solution exists for any placement of red/blue points still needs to be established. Determining the minimal number of intersection points to visit to cover all red/blue pairs is also not trivial. Therefore, a relationship exists between selecting a subset of intersection points for completeness and finding the shortest scenic route. Properly formulating this relationship in at least specific red/blue points configurations will help us comprehend the larger problem. One need not use red/blue as distinct colored points but use specific landmarks and create scenic routes that provide a scenic view of the landmarks, as shown for the capitol hill area in Washington DC as shown in Figure 5. A longer paper about scenic routes in 2D is in [RTK 2023].

## 3 Scenic Routes over Weighted Points in $R^2$ (2D) Space

In each 2D space, we can have points with different levels of importance. One would prefer viewing those points from a closer/farther position by their level of importance. A point in 2D where the user can view two given points by his/her preference of distance is termed a scenic point. We develop the concept of scenic paths in a 2D space for two points that have weights associated with them. Subsequently, we propose algorithms to generate scenic routes a traveler can take, which cater to certain principles which define the scenic routes.

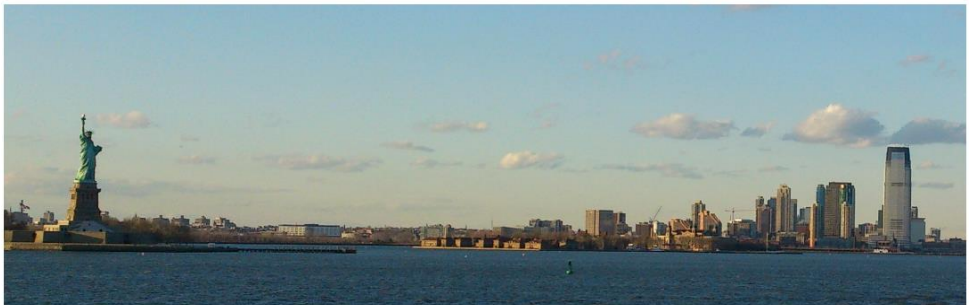
Figure 6: A scenic image of the Statue of Liberty with the 30 Hudson Street building (the tallest tower on the right among other skyscrapers [Simon 2014], licensed under CC2.0

Consider a rectangular area with *n* red and *m* blue points, each with its weight. The apparent weight of a given weighted point $P_1$ to another point $P_2$ is the ratio of the weight of point $P_1$ (say $W_{P1}$) to the distance between the two points, i.e., the apparent weight is equal to $W_{P1}/|P_1P_2|$.

A scenic point is where the apparent weights of a particular red and blue point are equal. Hence, the scenic point corresponding to a particular pair of red-blue points satisfies the following condition for a red point and a particular blue point (see Figure 7a):

$$w_1 d_2 = w_2 d_1$$

where $w_1$ and $w_2$ are the weights of the red and blue points, respectively ($w_1 > 0$, $w_2 > 0$, and $d_1$ and $d_2$ are the distances of the scenic point from the red point and the blue point, respectively). A path on which each point is a scenic point (in this case, $w_2 < w_1$) scenic is termed a scenic path, i.e., the locus of the point *P* in the above diagram would result in a scenic path.

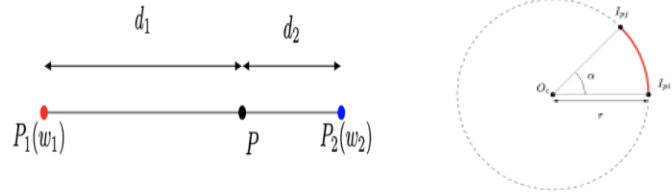

Figure 7: (a) A scenic point P (in this case, $w_1 < w_2$) [left image], (b) An arc of a circle *c* (with center $O_c$) with its endpoints at $I_{pi}$ and $I_{pj}$ [right image]

### 3.1 Formulation of the problem

We have a set of red points *R*, and a set of blue points *B*. Each point *p* can be defined by its coordinates *(x, y)* and the weight $w_p$ of the point. Let us say that the paths generated using the formula $w_{ri}\, d_{bj} = w_{bj}\, d_{ri}$ for all combinations of $(r_i, b_j)$, $r_i \in R$, $b_j \in B$, is *C*. A scenic path is either a circle (for the curve generated from applying the scenic path condition on two points with unequal weights) or a line (on two points with the same weight).

A scenic path is partitioned into scenic edges through its intersection with other scenic paths. This set of partitioned edges that belong to a particular path $c \in C$ is $E_c$. The weight of each edge is the length of the edge.

The length of an arc between two consecutive points $I_{Pi}$ and $I_{Pj}$ (see Figure 7b) on the same circle *c* is $r.\alpha$, where *r* is the radius of the circle *c* and $\alpha$ is the angle $\angle I_{Pi} O_c I_{Pj}$, where $O_c$ is the center of the circle *c*.

Let us say that the set intersection point formed due to the intersection of the curves is $I_P$. The edges between the intersection points of all such paths in *C* form the edge set $E_P$, i.e., $[c \in C\ E_c = E_P$. The graph over all such intersection points $I_P$ is $G(I_P, E_P)$.

The algorithms take $G(I_P, E_P)$ as input and output a scenic route, which is a subgraph of $G(I_P, E_P)$. Let the nodes in such a route be $S \subseteq I_P$, and the edges within the route be $E \subseteq E_P$.

Scenic routes have the following requirements, and the algorithms devised in this paper prioritize the requirements in the following order:

**Completeness:** Travelling on the route provided by the algorithm must allow one to view many red-blue pairs. It is desirable that a route would give a view of all red-blue pairs (i.e., all |R|.|B| pairs). The ability to view a larger number of red-blue pairs on a scenic route would add to the scenic beauty the route offers.

**Only scenic:** The route must consist of only scenic edges.

**Minimal route length:** The total route length i.e., the sum of all edges that are part of the route should be as low as possible.

**Minimal repeated edges:** A route must contain minimal repeated edges. Repeated edges are defined as stretches of paths that must be traveled more than once (repeated) to complete the entire route.

**The minimal number of edges:** A route must not have many edges. Traveling on a route must allow for long, uninterrupted stretches of scenic points. In other words, a route should have few direction changes.

These are paths whose one endpoint is free (not connected to any other edge, i.e. the free endpoint has a degree Repeated edges come into play to produce a closed path to return to the starting point.

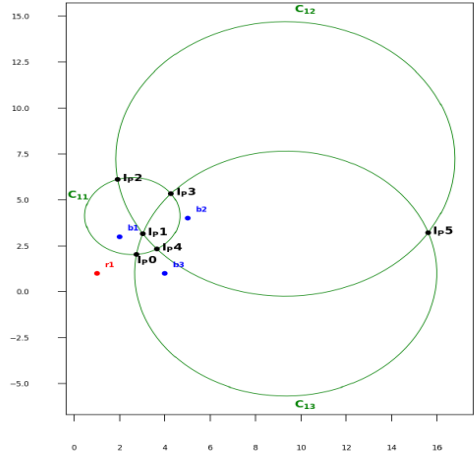

Figure 8: A sample graph with one red p*oin*t, and three blue points, the corresponding intersection points, and the scenic paths (points $r_1$, $b_1$, $b_2$, $b_3$ have weights 2.5, 1.5, 1.8, and 2)

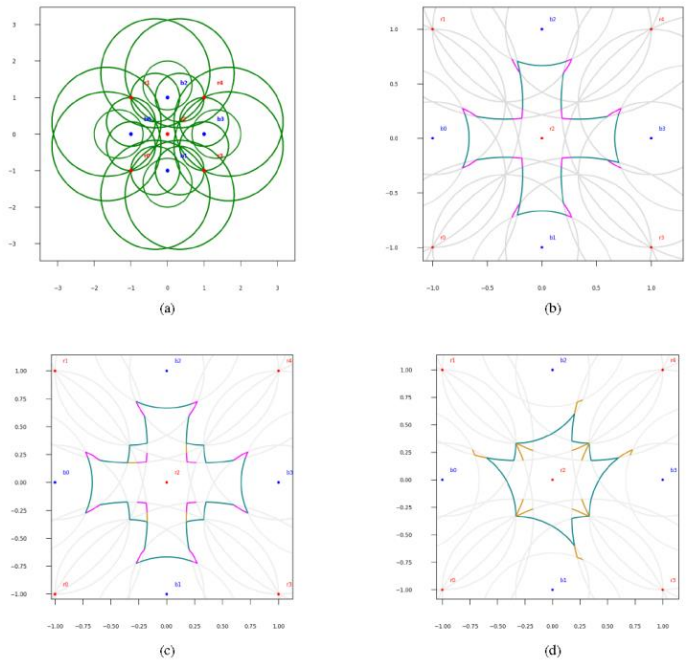

Figure 9: (a) Points organized in a grid, alternating red and blue points, and the resulting scenic graph. (b) Route generated by the ACU algorithm on the graph in (a). (c) Route generated by the ACCH algorithm on the graph in (a). (d) Route generated by the DPE algorithm on the graph in (a).

## 3.2 Algorithms for Scenic Routes with Weighted Points

The results of these algorithms are shown for a specific case in Figure 9. More results are in [STK 2023].

**All Curve Umbrella (ACU) Algorithm** bases their route on the shortest edge from each scenic path (i.e., the smallest edge from each $E_c$, $c \in C$). Since the endpoints of each edge are common intersection points between two or more curves, we are trying to enable the traveler to switch between scenic views with minimal traveling distance easily. This algorithm explains how these edges are connected to form a route.

**All Curve Convex Hull (ACCH) Algorithm** works on the Convex Hull [Jarvis 1977] of the set S rather than the endpoints of the edges in E. We connect the points of the hull with the APSP-specified path. While connecting the points with the edges in the APSP, we check whether previously inserted edges have already connected the two endpoints of each edge.

**Dense Point Expansion (DPE) Algorithm** aims to achieve completeness by choosing edges attached to well-connected points, i.e., points with a higher degree. A point with a high degree indicates that it would act as a hub between edges belonging to multiple scenic edges/circles, giving the traveler multiple path choices at every node. And hence, by choosing such points, we attempt to cover all possible scenic views with as minimal hops as possible, eliminating trivial edges in the route.

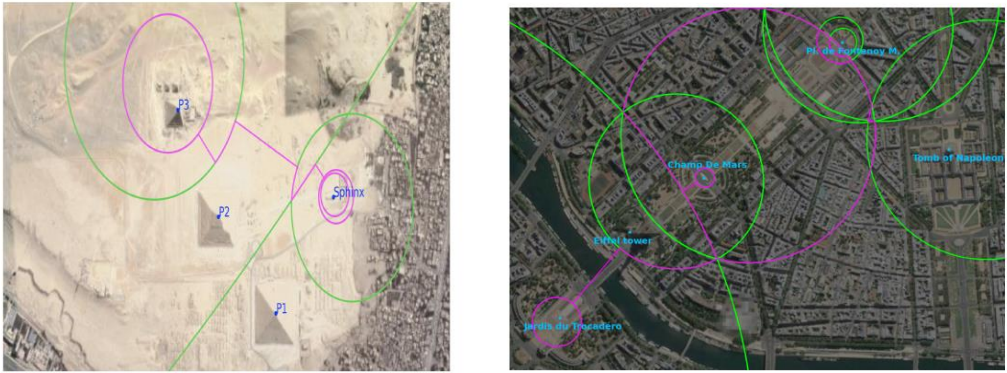

Figure 14: A scenic route (in pink color) on the area of the Great Pyramids (left), and the area around the Eiffel Tower (right), generated using the ACU algorithm.

**Examples:**

**Consider the Pyramids of Giza**, the four points of interest are "The Pyramid of Khufu" (P1), "The Pyramid of Khafre" (P2), "The Pyramid of Menkure" (P3), and "The Great Spinx" (P4). The weights of these points are set to their actual heights (P1: 138.5m, P2: 136.4m, P3: 61 m, P4: 20m). The ACU algorithm generated the scenic route (left image of Figure 14), the green paths (curves) are also scenic but not the generated scenic route. Note that since the heights of P1 and P2 are almost the same, the scenic path between them is almost a straight line.

**Consider the area around Eiffel Tower**, the five points of interest are "The Eiffel Tower", "The Jardis du Trocadero", "The Champ De Mars", "The Tomb of Napoleon Bonaparte", and the monument at the "Place de Fontenoy square". Their respective heights are 330m, 70m, 20m, 107m, and 16 m. The scenic route is shown in the right image of Figure 14.

## 3.3 Summary

The scenic route for weighted points is an interesting extension of the scenic routes in 2D problem. As seen in the examples above, some roads and landscaping overlap with the scenic routes; thus, these intersections provide a satisfying view of the monuments from these routes. In general, with the scenic routes over weighted points, the issues of establishing the existence of an optimal solution, getting the optimal solution, and ensuring the algorithm's complexity are to be addressed. The optimal criteria can be completeness (see all red/blue point pairs) and the shortest distance route. Some scenic curves may not intersect, and non-scenic segments must connect the scenic paths. Thus, minimizing non-scenic route length in the traversal to provide for the complete scenic route is challenging. A longer paper with details of scenic routes over weighted points in 2D is in [STK 2023].

## 4 Scenic Routes over $R^3$ (3D) Space

We can extend the work to 3D space where each red and each blue point are in XYZ space [Ravuri 2023]. Thus, based on a scenic point defined as an equidistant point between one red and one blue point, the bisecting plane between the two points is the two-dimensional scenic plane between the points. Two scenic planes can intersect to give a scenic line, and two scenic lines can intersect to give a scenic point. Thus, the scenic route will consist of traversal on a scenic plane to reach a scenic line and then to a scenic point to determine the scenic route. But these planes and lines may intersect near infinity or a long distance, and finding a scenic route within a bounded region requires some non-scenic path traversals. The problem in 3D is challenging, and establishing completeness and bounded distance traversal are aspects to study.

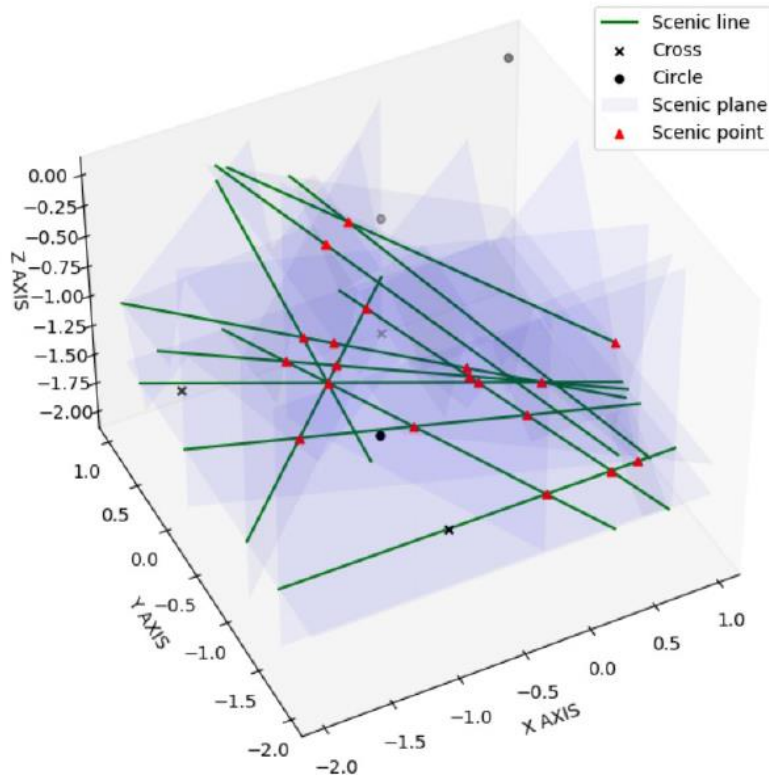

Figure 15: Three dots (•) and Three crosses (x) with scenic planes, lines, and points.

In Figure 15, there are three dots (•, red points) and three crosses (x, blue points) with equidistant scenic planes, the scenic lines where scenic planes intersect, and scenic points where scenic lines intersect. This orientation of the planes, lines, and points is mapped to a graph over which scenic route traversal can happen. The mapping from the physical manifestation of the geometric structures, and their intersection to a traversal over the weighted graph is a formalism that helps pursue the scenic route problem in 3D. The computational geometry aspects regarding planes, bisectors, and their intersections, along with their geometric properties, will guide the process of mapping the geometry of scenic aspects to the corresponding traversal graph. The graph itself could be disconnected and needs to be connected using non-scenic paths. The location and length of these non-scenic paths will require careful formalism to optimize the total scenic route traversal length. Thus, this work opens a new area of interest to practitioners and theory researchers. The scenic routes in 3D could be regions and lines where the drones can travel to provide scenic views and be equally separated from other red/blue points.

## 5   Scenic Routes over $R^d$

In the previous sections, we developed the notion of scenic points, lines, and planes between two points in $R^2$ and $R^3$. Our definition of a scenic area (a plane in the case of $R^3$) pertains to an equidistant area between two points in the case of greater than three dimensions. What would be the notion of the scenic area, and what aspects will arise if we consider the same notion of equidistant scenic areas and their intersections to generate the traversal graph?

Given a set of red and blue points in $R^d$, the scenic area between red and blue points will be a *d-dimensional* hyperplane that bisects them. Two hyperplanes can intersect to give a *(d-1)* hyperplane that provides a scenic view of two pairs of red and blue points. Progressively, the *(d-1)* hyperplanes can intersect to provide *(d-2)* hyperplanes, going to a scenic 3D plane, a scenic line, and a scenic point. Each such hyperplane will be a node in a traversal graph, and an edge exists if two hyperplanes (of any dimension) intersect. A traversal graph constructed will need node weights that give a path of minimal length to go from one hyperplane to another hyperplane through a third hyperplane. Since a node representing a hyperplane can connect multiple hyperplanes, the node weight is a vector giving the shortest path length connecting such connected hyperplanes at this node. Further, the traversal graph itself may be disconnected. In this case, we need not scenic paths connecting these disconnect traversal subgraphs.

Once the traversal graph is constructed, the densest hyperplane algorithm can provide a scenic route in $R^d$. That is, we consider the hyperplane connected to most hyperplanes as starting point of the traversal and then move to other nodes with fewer connections to hyperplanes. This algorithm will generate a connected scenic route if the traversal graph is connected. Otherwise, we will have partial scenic routes in each connected subgraph, with these partial scenic routes connected by not scenic paths. Thus, at least one algorithm determines a scenic route in $R^d$. Other algorithms with different strategies and properties can cater to different use case scenarios.

## 6   Nature of Scenic Point, Scenic Route, and Applications

 In this work, we stated a scenic point if it is equidistant to one red point and one blue point. This definition of a scenic point is simplistic yet powerful as it exposes the issues of the development of scenic route algorithms. The key takeaway from this formalism is mapping the geometric aspects of scenic areas and their intersections (connectivity) to a traversal graph. Thus, any other definition of a scenic point (say, based on the weight of the points) should also provide a similar mapping from geometric aspects to the traversal graph. The mapping is not trivial and has to provide suitable node attributes and edge attributes to capture the geometry aspects of scenic regions (areas) intersection and traversal between the scenic regions. Once this is correctly done, the scenic route extraction algorithms can be designed and evaluated.

One of the main applications of the two-color scenic route problem is that the points could represent points of two classes in a supervised learning problem, and the scenic route is a partitioning hyperplane between the two classes that show how the points of the classes are geometrically located in $R^d$ and the regions between them separated. Thus, it is possible to look at the geometric landscape of the points to comprehend the location of the points from the two classes. If required, the geometric landscape for up to three dimensions can be visualized based on dimensionality reduction techniques, otherwise, in a representative *2D* or *3D* data subspace. But for higher dimensions, some other techniques can be more suitable and need a redefined scenic point.

In our earlier work [VK 2009], we used a data visualization technique to determine the subspaces of $R^d$, in which one cluster's points are closer (within a *k*-nearest neighbor) to another cluster's points. Now that idea can be extended to find points of a class closer to points of another class across various subspaces. Now a scenic point can be a point that is a centroid of these two sets of points as it provides a comprehensive view of points belonging to both classes.

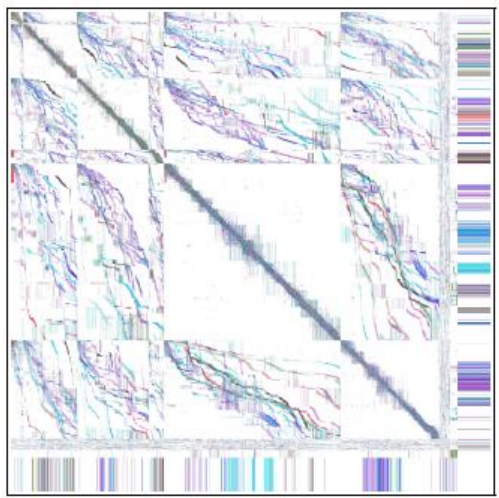

Figure 16: A Heidi image of 25-dimensional real estate data set.

In Figure 16, we have 46 classes, four of which are very large, and the rest are small. The Heidi visualization [VK 2009] of the data set is in Figure 16. Multiple subspaces overlap between points of two classes. These are single-color patterns among points of two different classes. Once we pick a pattern (the wavy single-color pattern), there will be points of the row class closer to points of the column class across multiple subspaces. Let row class points be $R_P$, and the column set of points be $C_P$; then the scenic point will be the centroid of points $R_P \cup C_P$, now since each row class has many patterns with each of the column classes, and vice versa, there will be multiple such connected centroids across all the classes. The traversal graph using these centroids can provide a scenic route across the classes, with the details of points visible from each centroid. The movement from the centroid (centroid jumping route) of one set of class points to another can provide intuition about the way is spread out in 25-dimensional space. Further insightful techniques can be developed to provide scenic routes and their traversal for higher dimensional data sets.

In the 2D and 3D cases, we considered only red/blue points, but in general, there can be many different colored points; in this case, the scenic point can be equidistant to any two-colored points; thus, the solution provided earlier can work on two-color restricted scenic route traversal.

Thus, as we move on the scenic route, we see that two colored points are equidistant. The notion of completeness would consider all pairs of differently colored points. Designing a provable optimal algorithm for generating scenic routes is challenging. One can consider three color scenic route problem, where from a scenic point, the three differently colored 2D points are equidistant, and the scenic route will include hopping from one scenic point to another across paths that provide scenic views of two differently colored points. In higher dimensions, providing scenic routes over points of many colors is challenging in defining the notion of a scenic point and setting the criteria that a scenic route must meet to design efficient algorithms to generate scenic routes. We expect the problem to be addressed as it is challenging and has applications in data visualisation, route planning, and landscape design.

# 7   Conclusion

In this work, we introduce the problem of scenic routes among points in $R^d$. The key development in this work is the nature of the problem in terms of both defining the concept of scenic points and scenic routes and then coming up with algorithms that meet different criteria for the generated scenic routes. The scenic routes problem provides a visual trajectory for a user to comprehend the layout of high-dimensional points. The nature of this trajectory and the visual layout of the points have applications in comprehending the results of machine learning supervised and unsupervised learning techniques. The red/blue points in our examples could be to be in a class or not to be in a class. The applications could include landscape design to adhere to the scenic beauty of the artifacts. The generation of equally separated layouts for designing composite hardware where interference could be an issue.

The nature of the scenic point in this work is kept as being equally distant from the two points. This characterization of the problem explodes in the order of $n^4$ scenic lines in 2D for n points. For 100 points, 100 million scenic lines are there; hence selecting appropriate lines or limiting points to representative points will provide an initial solution. Locally scenic routes can be generated to develop a scenic route network hierarchically. One can come up with different characterizations of the scenic point as a relationship of a point to a set of points (like centroid) and develop solutions for scenic routes.

The scenic route problem challenges theorists in terms of bound for the distance traveled for completeness or generating a compact scenic route with greater coverage and for applied scientists to explore applications of scenic routes in various domains.

**Acknowledgments:** This work is an overview of collaborative work done with Loay Rashid, Vijayraj Shanmugaraj, Lini Thomas and Yashaswini Ravuri.